# Homo economicus to model human behavior is ethically doubtful and mathematically inconsistent

Marc Lunkenheimer, University of Applied Sciences Neu-Ulm, Germany
Alexander Kracklauer, University of Applied Sciences Neu-Ulm, Germany
Galiya Klinkova, University of Applied Sciences Neu-Ulm, Germany
Michael Grabinski, University of Applied Sciences Neu-Ulm, Germany

**ABSTRACT**

In many models in economics or business a dominantly self-interested homo economicus is assumed. Unfortunately (or fortunately), humans are in general not homines economici as e.g. the ultimatum game shows. This leads to the fact that all these models are at least doubtful. Moreover, economists started to set a quantitative value for the feeling of social justice, altruism, or envy and the like to execute utilitarian calculation. Besides being ethically doubtful, it delivers an explanation in hindsight with little predicting power. We use examples from game theory to show its arbitrariness. It is even possible that a stable Nash equilibrium can be calculated while it does not exist at all, due to the wide differences in human values. Finally, we show that assigned numbers for envy or altruism and the like do not build a field (in a mathematical sense). As there is no homomorphism to real numbers or a subset of it, any calculation is generally invalid or arbitrary. There is no (easy) way to fix the problem. One has to go back to ethical concepts like the categorical imperative or use at most semi quantitative approaches like considering knaves and knights. Mathematically one can only speculate whether e.g. surreal numbers can make ethics calculable.

1. INTRODUCTION

In economics and many business situations, it is assumed that humans tend to look for maximal economic wellbeing. The theoretical model behind this is called homo economicus. However, numerous experiments have shown that humans are not homines oeconomici. The best-known test is the ultimatum game. In it player A gets goods (e.g. € 10) and is allowed to share it with B within any ratio (from € 9 for A and € 1 for B to € 1 for A and € 9 for B). If B accepts it, each player gets the share determined by A. If B refuses it, both will get nothing. If at least B were a homo economicus, he should always accept. However, in reality reaching a certain level of injustice B will not accept any more because many people perceive an unequal share between two equal individuals as unfair or unjust. The exact level when an individual perceives a share as unfair varies not only from person to person but is also different over time and will depend on e.g. whether A as a person is known. This is already a first hint that *justice* cannot be quantified in a mathematical sense and that certain values (such as justice) might affect human behavior and lead to non-rational decisions in context of the homo economicus model.

Nonetheless, many decisions in business are made based on the model of homo economicus. In business administration, all decision models, optimization calculations, or variable salary models depending on performance are good examples. Let us assume two companies A and B facing a decision whether they start to invest abroad or not (Figure 1). If both companies do not invest abroad, they can both expect a profit of 4 each. In the same way, both will get a profit of 2 if both are investing abroad. A mathematician will immediately recognize that staying at home is not only maximizing profit, it is also the only stable Nash equilibrium. Any move from there will reduce the individual profit. Reality tells often a different story. Let us further assume going from (4,4) to (3,2) or (2,3) will increase the *reputation* of A or B, respectively. This leads to a deviation from the more profitable position due to certain character traits, such as envy of the other's reputation. Envy could obviously be so strong that A and B are willing to pay for it, ending up with half the profit in the end.

|  | company A | |
|---|---|---|
| company B | stays home | invests abroad |
| stays home | (4,4) | (3,2) |
| invests abroad | (2,3) | (2,2) |

**Figure 1.** Payoff matrix when staying home or investing abroad with (profit of A, profit of B)

According to this example, the players' behaviour indicates that either they are acting irrationally or it is not explainable by the homo economicus model. One explanation would be an accounting for envy, altruism, and the like. In Figure 1 companies are willing to half their profits because of envy. If envy is worth half the profit, we have a number for envy. One may include it in the payoff matrix of Figure 1 leading to a stable Nash equilibrium at (2,2). In doing so one gets the result from reality though strict homines economici are assumed. This is often done in modern economics though it is extremely flawed. There are many other examples where one is misled if one assumes a homo economicus in situations like this.

In this article, we aim to show that this application of homo economicus bears inherent mathematical contradictions that question the logical consistency of the model. Since an inconsistent set of statements can be used to prove anything, including nonsense, absence from contradiction is indispensable for useful scientific theories. If our argumentation were successful, the homo economicus model would be methodological useless for scientific reflection and modelling of human behavior with moral values because of its inconsistency.

After a short review on the current understanding of the problem in section 2, we will give a philosophical perspective on how economics started to explain every human (economic) behavior by maximizing the individual profit in Section 3. After that, the main point of section 4 is to show that quantification of ethics is not working, is not only a philosophical remark. We can rigorously prove that it is impossible for mathematical reasons too. In Section 5 we conclude that a quantification of ethics does not make sense, neither from a philosophical nor mathematical point of view.

## 2. LITERATURE REVIEW

The model of homo economicus has been challenged with experiments like the ultimatum game, as stated above (Dilek & Kesgingöz, 2019). In a recently published article, Schreck, van Aaken and Homann (2020) have therefore defended homo economicus in the context of social dilemma situations against criticism, especially from the empirical side. The authors put forward an argument that sets the contradictory empirical observations on the homo economicus model analog to those on Galileo Galilei's law of motion: "Interpreting observed moral behavior as a falsification of the homo economicus model is like interpreting the slow fall of a feather as a falsification of Galileo's laws of motion." (ibd.) Instead, the discrepancy between predicted behavior and actually observed behavior should be investigated. They suggest to see models of moral behavior that are based on social preferences not as alternatives but refinements of the homo economicus model. In their view, the purpose of these approaches is to contribute to a contingency theory of moral behavior, rather than falsifying and preplacing the homo economics model.

The argumentation of Schreck et al. with regard to the scientific-theoretical foundations seems convincing at first glance. However, the aim of our publication on this is not to repeat empirical counter-arguments against homo economicus, but to reflect on the model with regard to its methodological foundations and the mathematical assumptions made in the process for quantifying qualitative values.

### 3. THE PHILOSOPHICAL POINT OF VIEW

Before deep diving into mathematics, a philosophical perspective aims to broaden the understanding of the main problem here. Moral philosophy, or ethics, is the methodical reflection on human action and under what conditions it can be called "morally right" or "good". Various ethical theories answer this question differently. Utilitarianism is a form of ethics that, reduced to its essentials, states that an action is morally right precisely when it maximizes the aggregate total utility.

Jeremy Bentham, founder of systematic utilitarianism, put this in the formula: "the greatest happiness for the greatest number" (Bentham, 1789). This principle states: "By the principle of utility is to be understood that principle which universally approves or disapproves of every action in proportion as it appears to have an inherent tendency to increase or diminish the happiness of the group whose interest is in question." (ibd.) The measurement of happiness is determined as follows: "Add up the values of all the pleasures on one side and all the sufferings on the other. If the side of pleasure predominates, the tendency of the action with regard to the interests of that individual person is good on the whole; if the side of suffering predominates, its tendency is bad on the whole." (ibd.) This definition of the utilitarian approach to ethics exposes a natural proximity to the homo economicus model. In order to maximize happiness, he describes what economists call a cost-benefit reasoning. Thereby, this normative theory could be used to open up an ethical perspective upon business related issues, since economic thinking is inherent.

But can values such as happiness, justice, or even human life be quantified and aggregated with arithmetic operators? Critics have challenged this approach (Taylor, 2017). They state that, moral values cannot be quantified, measured, or even calculated. To evaluate actions as morally "better" or "worse" in a quantitative sense seems to be a linguistic convention alone, but does not sustain ethical reflection.

An example illustrates very well what the problem is when qualitative values are quantified. In the 1960s, Japanese and German cars in the subcompact car class increasingly dominated the US car market. In 1968, Ford executives decided to launch the Pinto to challenge the competitors. A very tight schedule was to be followed to start with an early production of the new car in order to eliminate the foreign invaders from the market as quickly as possible.

Before Ford started producing their new car, they tested several prototypes. The crash-tests showed that the Pinto posed a serious danger to its occupants, as it easily caught fire when involved in a rear-end collision. Ford officials faced a decision. Should they stick to the current design and thus ensure that the ambitious schedule can be met, thereby putting the safety of its customers at risk? Or should they suspend the planned start of production and give the tank a redesign to make it safer which would result in the home market being dominated by foreign companies for another year? (Lineq, 2001)

The evidence suggests that Ford relied, at least in part, on cost-benefit reasoning. Hereby, Ford concluded that, if they agreed on the technical upgrade of the tank design, the increased cost outweighed the benefit. Interesting is the calculation they made. They estimated the technical improvements to prevent gas tanks from leaking and prevent fire hazards to be $11 per vehicle in additional cost. They planned with a production of 12.5 million cars. This results in $137 million total additional cost if they would upgrade the gas tank. But what would be the benefit from that? On the benefit-side (if the gas tanks were replaced and the inmates life were saved), they estimated $49.5 million. They did the following calculation: They assumed society loses $200,725 every time a person is killed in a car accident. They broke down the cost in various positions such as direct and indirect productivity losses ($132,000 + $41,300), hospital ($700), insurance administration ($4,700), legal and court expenses ($3,000), employer losses ($1,000), property damage ($1,500), victim's pain and suffering ($10,000), funeral ($900), lost consumption (5,000$), miscellaneous accident costs ($200) and other ($425). They estimated 180 burn deaths ($200,000 per death), 180 serious injuries ($67,000 per injury), and 2,100 burned vehicles ($700 per vehicle) which leads to a total of $49.5 million. Since the cost of a new tank design ($137 million) clearly

outweighs the benefit ($49.5 million), Ford's management decided to stick with the original design und not upgrade the Pinto's fuel tank, despite the results reported by its engineers. Between 1971 and 1978, the Pinto was responsible for a number of fire-related deaths. Ford puts the number at 23; its critics say the figure is closer to 500. For the details on the Pinto case, see Lineq (2001).

The simple example shows the problem that arises when values are quantified: their quantity remains arbitrary. Why do these individual positions determine exactly a loss of a human life? Why is only the suffering of the person directly involved considered, but not that of his or her relatives? Why is individual suffering and pain worth exactly $10,000 and not more or even less? Economists tend to neglect this normative foundation of their argumentation or if they do, it remains arbitrary.

## 4. ACCOUNTING FOR ETHICS IN ECONOMICS

Considering ethics, morals, and many other values is necessary in economics. A good example is the theory of knaves (self-interest only) and knights (common-good oriented) as explained from Grand (2003). This theory suggests that not every human behavior is driven from the same motivation. Especially when it comes to salary, the output of a person's work depends not only from the amount of money she or he earns. Whether the person is a knave or a knight is equally important. For example, it has been applied to a new pay scale for German professors introduced in the beginning of this century (Grabinski, 2005). The main point was changing a system based on a fixed salary plus some growth due to gain of experience over time into a base salary plus some bonus depending on output quality bargained every year. In Figure 2, the dashed curve denoted by $S_{kna}$ shows the typical behavior of a self-interested knave. Output and payment are about linearly related. The solid curve denoted by $S_{kni}$ stands for a typical knight. For a rigorous theoretical discussion, please see Grand (2003). The points *A* and *B* state the payment and output quality of a typical knight or knave, respectively. As one clearly sees, a knight has a higher output with lower pay while it is vice versa with the knave. Besides it is very difficult to measure the output quality of professorial work, we have a much more severe problem. Professor are supposed to be dedicated to their work.

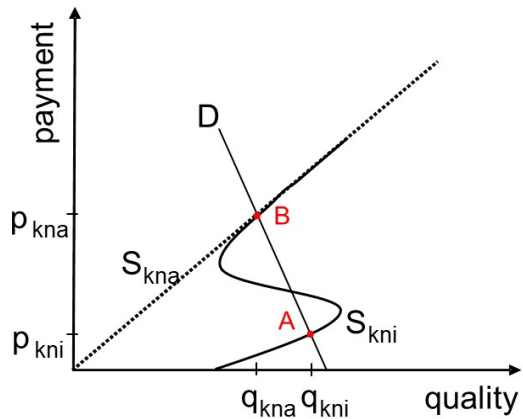

**Figure 2.** Curves of payment over output quality from Grabinski (2007).

This is nicely expressed in the German word *Berufung* used for the appointment of a professor. It is the same word as used for a priest being *appointed* by god. So a dedicated professor should have another motivation besides pay. However, treating a knight like a knave by trying to control him by varying the payment, will lead to a phasing out denoted by the line D in Figure 2. We end up with higher cost and lower quality due to assuming self-interest as the main trait.

In Figure 2 the quantitative deviation from home oeconomicus has been shown. Please note that this does not show how to *calculate* the behavior of Mother Teresa who was for sure an extreme knight. The whole consideration is semi quantitative only. In Figure 2 one can only show the (negative) effect at work. But the exact curves (especially $S_{kni}$) can not be constructed. (Of course the gaps in Figure 2 are for sure exaggerated) Trying to construct the lines by e.g. differential equation would be mathematical nonsense as there is no homeomorphism from the space of dedication, curiosity, etc. to the $\mathbb{R}^n$. But this will be discussed in more detail in Section 5. The next example shows such misuse of numbers in a simple example.

Now we come back to the game of Figure 1 in the introduction. Mathematically it is supposed to end in (4,4). Due to envy (2,2) is a typical end in reality. As stated envy makes A in point (3,2) feel better and B worse. So one may come up with the idea to include some envy parameters in Figure 1. In doing so we end up with Figure. The parameters $\alpha$ to $\delta$ denote the envy. So if A has a profit of 3 but B a profit of 2, A *feels* a profit of $3 + \beta$ and B is so annoyed as were its profit $2 - \beta$. With $\alpha = \beta = \gamma = \delta = 1.5$ the problem appears to be solved. The point (2,2) is a stable Nash equilibrium showing unfortunately the lowest total profit. It is like in the game prisoner's dilemma. The effect of envy even appears to be calculable. Every $\alpha, \gamma > 1$ and $\beta, \delta > 0$ will create a strong enough envy effect to end up in (2,2) while $\alpha, \gamma \leq 1$ will lead to (4,4) as a Nash equilibrium.

Nevertheless, Figure 3 gives an explanation in hindsight. In order to predict the future one has to know the envy parameters $\alpha$ to $\delta$ beforehand. In order to do so, one needs a rigorous definition envy. This is maybe possible by certain hormone levels. Then we need a homomorphism from these parameters to $\mathbb{R}$. It will be impossible. If A envies B *and* C by a certain amount, will the hormone levels add up? Will this lead to simply added envy in our parameters $\alpha$ to $\delta$? Furthermore, the amount of envy will not only depend on the difference of profit in Figure 1 but whether A likes or hates B. Mathematically spoken we have no transitivity.

|  | | company A | |
|---|---|---|---|
|  | | stays home | invests abroad |
| company B | stays home | (4,4) | $(3 + \alpha, 2 - \beta)$ |
|  | invests abroad | $(2 - \gamma, 3 + \delta)$ | (2,2) |

**Figure 3.** Payoff matrix including envy; (Profit A; Profit B)

One could also say that there are many more variables. But this makes it worse. We would need a homomorphism to $\mathbb{R}^n$ for all these variables. Even in the very unlikely situation that one can find a homomorphism, one will probably end up in a chaotic system. This is very similar to a stock market. The price of a particular stock is determined by each individuum's willingness (for whatever reason) to buy or sell a particular stock at a particular price. Though this sounds like a simple predicting rule, it leads to an unpredictable chaotic stock market (Appel & Grabinski, 2011; Klinkova & Grabinski, 2017). So one might at least say that one can make a rough estimate. Due to chaos this rough estimate may be very rough. Having a very wide range for e.g. envy parameters and the like, creates another problem as has been shown for game theory in a very recent publication (Klinkova & Grabinski, 2021).

In Figure 4 a game has been introduced where each player has a choice of three strategies. It could be applied to two competitors trying to conquer the market with three strategies. In (Klinkova & Grabinski, 2021) the fun example of penalty shooting in soccer has been scrutinized. The following payoff matrix Figure 4 has been taken from this (ibd.) publication where further details can be found. Estimating the $p_{ij}$ from real championship games (The Economis, 2018), it has been calculated in (Klinkova & Grabinski, 2021) that jumping and shooting with 43.0 % probability to the left and with 37.5 % to the right (and 19.5 % else) would be the optimal strategy for player and keeper. As calculated in (Klinkova & Grabinski, 2021) it will lead to a probability for a goal of 79.9 %.

|  |  | player | | |
|---|---|---|---|---|
|  |  | left | middle | right |
| keeper | left | $(p_{ll}, 1 - p_{ll})$ | $(p_{lm}, 1 - p_{lm})$ | $(p_{lr}, 1 - p_{lr})$ |
|  | straight | $(p_{sl}, 1 - p_{sl})$ | $(p_{sm}, 1 - p_{sm})$ | $(p_{sr}, 1 - p_{sr})$ |
|  | right | $(p_{rl}, 1 - p_{rl})$ | $(p_{rm}, 1 - p_{rm})$ | $(p_{rr}, 1 - p_{rr})$ |

**Figure 4.** Payoff matrix in penalty shooting; $p_{ij}$ are the success probabilities if keeper choose $i$ and player $j$.

Even in soccer there are other soft variable besides the success probabilities $p_{ij}$. Player or keeper may have stress factors because their career is in danger. But as stated, the $p_{ij}$ may also be success rates in investment strategies. So one may include envy like in Figure 3. As discussed above these envy parameters may have a broad distribution. Of course one can use the average only.

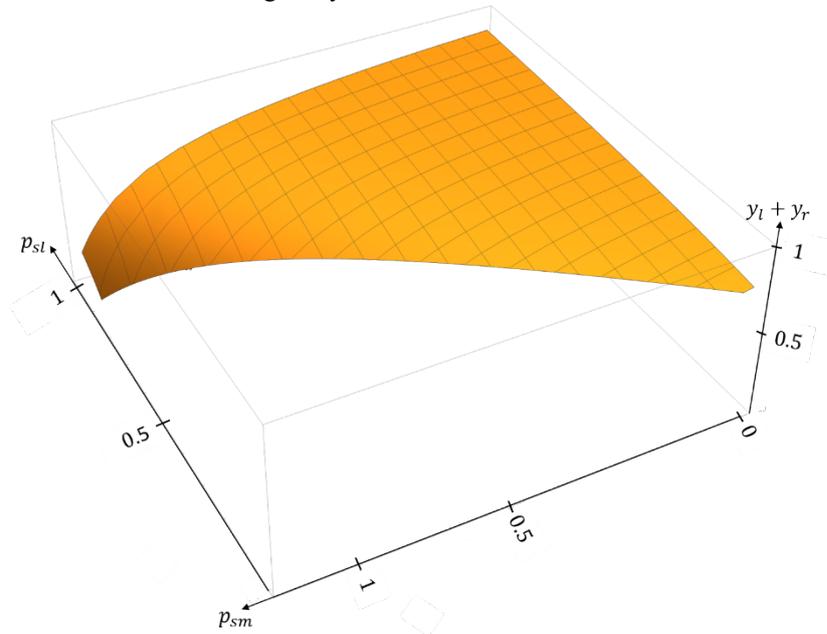

**Figure 5**. Regime of Nash equilibrium for varying $p_{sl}$ and $p_{sm}$

To see the point all $p_{ij}$ in Figure 4 may keep their original value but $p_{sl}$ and $p_{lm}$ are broadly influenced by envy. So they may vary between 0 and 1. Figure 5 has been directly taken from Klinkova & Grabinski (2021). As explained there, only in the orange meshed area there exists a Nash equilibrium. (In the area not covered one had to shoot e.g. with over 100 % probability) Even if the *average* values (including envy) for $p_{sl}$ and $p_{lm}$ lie the orange meshed area, the entire calculation does not make any sense. One gets a result, but it is a weighed average of an existing solution and a non-existing one. It is even quite dangerous. One may get a result and of course one relies on it, though it has no meaning. Please note that calculating the area in Figure 5 is straight forward within this example. But this is due to the particular payoff matrix of Figure 4. To find the orange meshed area like in Figure 5 is for an arbitrary case would require fundamental research in e.g. topology, which has been explained in Klinkova & Grabinski (2021).

## 5. MATHEMATICAL REQUIREMENTS TO *CALCULATE* ETHICS

From the examples from Section 4 it is clear what one needs. Objects like envy, greed, hate, altruism... must be measured in the first place. This could be extremely difficult, and mankind is far from being successful in this field. But even if done successfully one has to assign numbers to it in order to make any calculation. Consider the example of rating the most favored color for a new car. Firstly, one has to define the setup clearly. Does it mean a possible increase in price for a car being identical except for the color, the wellbeing while driving in such car or seeing such car. Going for the latter one may come up with detailed numbers. The color is uniquely defined by a wavelength (but only if it is a pure color, else one will have a distribution of wavelengths). The amount of happiness can be measured by hormone levels such as dopamine.

Now we would have numbers which are scientifically sound. An average wavelength is a color in between. But will the dopamine level do the same? If the dopamine level (in whatever units) is 50 while watching a red car and 10 while watching a yellow car, does it mean it should be $(50 + 10)/2 = 30$ when an orange car passes by? Instead of measuring dopamine levels one may ask people to rate colors on a scale from 1 to 5. This produces some arbitrariness but will show the same weakness.

Assigning numbers is often easy. But is an addition defined? How is it defined? It is a typical task for first year students of mathematics in linear algebra to find out whether some objects and a generalized addition (e.g. $\oplus$) will build a group. Unfortunately this is almost never done in business or economic research. From the examples of Section 4 the problems become obvious.

Dealing with numbers we need at least a reasonable defined addition and multiplication. Mathematically spoken the numbers must build a field. Probably everyone knows two fields: The real numbers $\mathbb{R}$ and the rational numbers $\mathbb{Q}$ with $\mathbb{Q} \subset \mathbb{R}$. As the first-year math students invent groups or fields with strange objects, one may say that one just needs creativity to define an addition to e.g. envy levels of Figure 3. Sadly, one can show that any (one-dimensional) field is homomorphic to $\mathbb{R}$ or a subset of it. If objects build a field, it is just a systematic renaming of real numbers. So there is no chance overcoming the problems of Section 3 when using real numbers or a homomorphism of it.

Leaving real numbers is also no trivial option. The well-known two-dimensional field of $\mathbb{C}$ has the drawback that it is not ordered. As there is no higher dimensional field and any two-dimensional field is isomorphic to $\mathbb{C}$, even the greatest creativity will not help. To see the point why order is essential let's go to the mathematical definition of a game: Given be a map $p$ with

$$p : S_1 \times S_2 \times \ldots \times S_n \to \mathbb{R}^n \qquad (1)$$

The $S_i$ are the possible strategies. $(\sigma_1, \ldots, \sigma_n) \in S_1 \times S_2 \times \ldots \times S_n$ be a particular strategy bundle. If for *any* $\sigma_i^* \in S_i$

$$p_i(\sigma_1, \ldots, \sigma_{i-1}, \sigma_i^*, \sigma_{i+1}, \ldots \sigma_n) \leq p_i(\sigma_1, \ldots, \sigma_{i-1}, \sigma_i, \sigma_{i+1}, \ldots \sigma_n), \qquad (2)$$

holds, then $(\sigma_1, \ldots, \sigma_n)$ is called Nash equilibrium. $p$ is the "profit" of a certain strategy. As one sees from inequality (2), one needs an ordered field. And of course everywhere in economics one wants e.g. to maximize a profit which demands order. Therefore mathematician always speak of real games in game theory.

Please note the problem of having no field is not limited to quantities like envy, greed, hate, altruism... In utility analysis one assigns e.g. points to potential profits when investing in A, B, or C. The profits itself or its estimates can be added. But the very point for assigning points is that such estimates are very hard to get. So the points are at most a monotonous function $f_i$ of the profit $p_i$. One wants to maximize

$$\sum_{i=1}^{N} p_i \;\to\; \max. \qquad (3)$$

But instead one maximizes

$$\sum_{i=1}^{N} f_i(p_i) \;\to\; \max. \qquad (4)$$

As $f_i$ is arbitrary and has the only restriction of being monotonous, the maximization in (3) and (4) will lead to arbitrarily different result depending on $f_i$. (Management consultants might know it when they are using utility analysis to *prove* a best strategy desired by themselves or the CEO) Only if the $f_i$ were linear functions, utility analysis could work. But then one has to guess the weights accurately which is far from being easy. The falsification of utility analysis has been shown in Grabinski (2007) already. There almost no hardcore mathematics has been used.

The learning from this section and in some sense of the whole publication is that in order to add or multiply numbers, addition and multiplication must be *defined* first. The statement sounds trivial and indeed it is. However assigning a number to an object sounds trivial and arbitrary. As one learns in primary school how to add and multiply, people are doing it without thinking. Nobody would say e.g. one can add *envy* to *hate*. But as soon these values are quantified, an addition is performed. However it is by no means better then defining

$$envy + hate \equiv envyhate \quad \wedge \quad \frac{envy + hate}{2} \equiv \text{eave} . \qquad (5)$$

## 6. CONCLUSIONS

We have shown that quantifying values like envy, greed, hate, altruism... may or may not be possible. But making calculations like finding optima with it, is almost ludicrous. Although, it appears to grow more and more common. We have also shown that there is no easy fix and maybe none at all. For the time being it is the best advice not using such *tools*. And even more importantly, not to rely on results created by such methods. Positively, a cost saving is possible by just not doing such calculations. We therefore conclude: The homo economicus model is neither a methodologically reasonable nor a normatively acceptable model for describing human behavior in business and economic contexts.

The point we made in this article may seem trivial from both a philosophical as well as from a mathematical point of view. However, the consequences for business administrations and economics are certainly not.

Leaving such evaluations to philosophers is one way out. For sure it will not please quite a few economists. But it will make economics more *human* and more scientific, as it is the basis of any science to find a proven answer. Peter Ulrich's Integrative Economic Ethics (Ulrich, 2008) is an approach for the use in economic context. This is based on the inviolable human dignity of every human being and should therefore not be forgotten in the context of economic action. According to Ulrich, a meaningful economy is one in which economists see the term "economic activity" as more than just acting for profit. Reasonable economic activity, in the sense of sensible and ethically correct, serves life, so to speak, since it integrates - in addition to survival through profit - the social and fulfilling aspect that society needs to flourish. The concept of knights and knaves (Grand, 2003) is another promising invention. Though it is hard to say who invented the concept, Richard Titmuss (1907 – 1973) was for sure an essential contributor.

To find a rigorous calculation of ethical things is at least a very long way to go. For sure it will look very different from what people expect. Real numbers ($\mathbb{R}$) could not be involved. So the result could not be 42. On the other hand, a similar situation happened at the beginning of the 20$^{th}$ century. Classical mechanics assigns any object a position and a velocity in order to describe it. These are all real numbers. Though classical mechanics is often an excellent approximation it is generally wrong. Fixing the problems is called quantum mechanics. Meanwhile it is easy to see (prove) that such theory is impossible to formulate with real number. One needs complex numbers $\mathbb{C}$. As stated above, complex numbers are not ordered. One cannot decide whether a number is bigger or smaller than another one. Therefore in quantum mechanics it is impossible to state e.g. whether an object is further to the left than another one. One can only speak of a certain probability of being further to the left.

Whether complex numbers can solve our problem here, can be answered with *maybe*. But it is a very big *maybe*. A may be better candidate are surreal numbers. In real numbers one says sometimes $a$ is arbitrarily close to $b$. However, for every $a < b$ there exists a $c$ with $a < c < b$. Admittingly over simplified, surreal number are real numbers plus infinitesimal close number. With these numbers it would be impossible to find a $c$ as defined above. John Conway (1937 - 2020) could prove that it is possible to build a group with them though no field. John Conway also worked on game theory though he generally considered games like go (圍棋) or chess. In these games there is of course also a finite number of strategies. But they are so big that even supercomputers are far from able to consider them all. Finding the best strategy needs something like creativity. John Conway observed some surprising connections between such games and surreal numbers (Conway, 1978). After Conway died of Covid-19 in 2020, research fortunately goes on (Henle, 2021).

**Author Contributions:** All the authors (ML, AK, GK, and MG) contributed to conceptualization, formal analysis, investigation, methodology, writing original draft, writing review and editing. All authors have read and agreed to the published form.

**Funding:** This research received no external funding.

**Institutional Review Board Statement:** Ethical review and approval were waived for this study because it does not involve any experiments with animals or humans.

**Informed Consent Statement:** This study does not involve any humans.

**Data Availability Statement:** This publication did not use any data not published within the paper or its references.

**Acknowledgments:** MG is grateful to Hans-Michael Ferdinand for discussions on selling indulgence being related to quantitative economics in e.g. game theory.

**Conflicts of Interest:** The authors declare no conflicts of interest.